\documentclass[twocolumn,preprintnumbers,amsmath,amssymb,superscriptaddress]{revtex4-1}
\usepackage{epsfig,bm}
\usepackage{ulem}
\usepackage{amsmath,color}

\begin{document}
\title{Total reaction cross section on a deuteron target and the eclipse effect
of the constituent neutron and proton}
\author{W. Horiuchi}
\affiliation{Department of Physics,
  Hokkaido University, Sapporo 060-0810, Japan}
\author{Y. Suzuki}
\affiliation{Department of Physics, Niigata University, Niigata 950-2181, Japan}
\affiliation{RIKEN Nishina Center, Wako 351-0198, Japan}
\author{T. Uesaka}
\affiliation{RIKEN Nishina Center, Wako 351-0198, Japan}
\affiliation{Department of Physics, Saitama University, Saitama 338-8570, Japan}
\author{M. Miwa}
\affiliation{RIKEN Nishina Center, Wako 351-0198, Japan}
\affiliation{Department of Physics, Saitama University, Saitama 338-8570, Japan}

\begin{abstract}
\begin{description}
\item[Background]
  Eclipse effect of the neutron and proton
  in a deuteron target is essential to correctly describe
  high-energy deuteron scattering. 
  The nucleus-deuteron scattering needs 
  information not only on the nucleus-proton
  but also the nucleus-neutron interaction,
  for which
no direct measurement of the nucleus-neutron
    cross sections is available for unstable nuclei.
\item[Purpose]
  We systematically evaluated the total reaction
  cross sections by a deuteron  target
  to explore the feasibility of extracting
  the nucleus-neutron interaction from measurable cross sections.
\item[Methods]
High-energy nucleus-deuteron collision is
described by the Glauber model, in which 
the proton and neutron configuration
of the deuteron is explicitly taken into account.
\item[Results]
  Our calculation reproduces available
experimental total reaction cross section data
on the nucleus-deuteron scattering.
The possibility of extracting the nucleus-neutron
total reaction cross section from nucleus-deuteron
and nucleus-proton total reaction cross sections is explored.
The total reaction cross sections of a nucleus by proton, 
neutron, and deuteron targets can be expressed, to good accuracy, in terms of 
the nuclear matter radius and neutron skin thickness.
Incident-energy dependence of the total reaction cross sections
is examined.
\item[Conclusions]
  The total reaction cross section on a deuteron
  target includes information on both the nucleus-neutron and 
   nucleus-proton profile functions.
  Measuring the cross sections by deuteron and proton targets
  is a promising tool to extract the nuclear size properties.
\end{description}
\end{abstract}
\maketitle

\section{Introduction}

Over half a century ago, Glauber and Franco examined 
high-energy antiproton-deuteron ($\bar{p}d$)
scattering and pointed out that the $\bar{p}d$ absorption 
cross section was always smaller than 
the sum of $\bar{p}p$ and $\bar{p}n$ cross sections~\cite{Glauber, Franco66}.
This cross-section defect was significantly large and its origin 
was explained by the so-called `eclipse' effect 
that the neutron and proton in the deuteron cast individual shadows.
When either particle lies in the shadow cast
by the other, it absorbs less effectively than when outside it.
As stressed there, 
the high-energy deuteron scattering is in fact not simply explained only
by the geometrical eclipse but
multiple scattering effects have to be taken into account. 
Total reaction cross section is a quantity mostly used to
measure absorption in nuclear collisions. 

The purpose of this paper is to study the total reaction cross section 
of a projectile nucleus $P$ scattered by a deuteron 
target with the beam energy of several tens MeV to a few GeV per nucleon, 
especially focusing on the eclipse effect, that is, the difference of the 
$Pd$ total reaction cross section from the sum of  total reaction 
cross sections of $Pn$ and $Pp$.  As the  effect is 
influenced by the size of $P$ and the property 
of  underlying nucleon-nucleon interaction, e.g.,
difference of the $pp$ and $pn$ total cross sections,
this  study is expected to be closely related
to the issue of radii of neutron and proton of $P$.

Measurements of interaction
and total reaction cross sections with the help of 
inverse kinematics have unveiled various exotic structure of
unstable nuclei, e.g., halo~\cite{Tanihata85,Tanihata13,Bagchi20},
neutron-skin structure~\cite{Suzuki95}, and recently the anomalous
growth of nuclear radius of Ca isotopes is found~\cite{Tanaka20}. Although 
such measurements have mostly been done on a carbon target, 
a deuteron target combined with a proton target appears to be superior in 
extracting the radii of proton and neutron. Furthermore, 
experiment on charge-changing cross sections
has recently been carried out to extract 
information on radius~\cite{Estrade14,Kanungo16},
but the experiment often 
poses some problems in both its 
data analysis and a theoretical formulation from the point of view of 
a reliable determination of the radius~\cite{Bhagwat04,Yamaguchi11,Terashima14,Suzuki16}.

Thanks to the recent development
of a proton target~\cite{Nishimura10,Moriguchi13,Moriguchi20},
the nucleus-proton interaction for unstable nuclei
can now be studied and gives valuable input 
to explore the property of neutron skin 
thickness, the difference in radii
between neutron and proton~\cite{Horiuchi14,Horiuchi16}. 
No such study is, however, available for the nucleus-neutron interaction.  
By understanding  quantitatively the eclipse effect of deuteron cross section, 
we expect to get information on the nucleus-neutron interaction. 

The deuteron has unique advantages that it has only one bound state, 
the ground state, and its wave function is readily calculated. 
The deuteron is fragile and can easily dissociate without inducing 
the mass-number change of $P$ in the collision. 
Such inelastic scattering process is difficult to identify in a 
measurement, and thus the interaction cross section is actually measured. 
A theory usually calculates the total reaction cross section.
We discuss the difference between  both the cross sections.

The paper is organized as follows.
In the next section, we present a formulation to describe
the high-energy nuclear collision.
Expressions for the $Pd$ total reaction cross section
as well as the interaction cross section
are derived within the Glauber model,
including the spatial correlations of proton and neutron in the deuteron.
Section~\ref{results.sec} shows our results.
First in Sec.~\ref{test.sec} we test the validity of
the present model by comparing
the theoretical total reaction cross sections on a deuteron target
with the available experimental data for known nuclei.
We quantify the deuteron eclipse effect
and investigate its magnitude systematically.
In Sec.~\ref{extract.sec}, we discuss the possibility
of extracting the nucleus-neutron interaction
from simultaneous measurement of the total reaction cross sections
on deuteron and proton targets by performing a numerical ``experiment''.
Two practical examples are given.
For future measurement, we show the universal behavior of
the total reaction cross sections
on deuteron, proton, and neutron targets
for a wide range of incident energy and mass number
using realistic density distributions.
Section~\ref{emp.sec} relates the total reaction cross sections
to the nuclear size properties, matter radius and neutron skin thickness.
Sensitivity of the cross sections to the radii of proton and neutron
is quantified through an investigation of their incident-energy dependence.
Section~\ref{interaction.sec}
compares the magnitude of the total reaction
and interaction cross sections.
The conclusion is given in Sec.~\ref{conclusion.sec}.

\section{Glauber model for deuteron cross sections} 
\label{model.sec}

\subsection{Total reaction cross section}

Here we employ the Glauber model to describe the high-energy
$Pd$ scattering. The validity of this approach was well tested
in $^{58}$Ni-$d$ scattering~\cite{Yabana92a, Yabana92b, Suzuki03}.
Let $\Psi^P_{\alpha}$ and $\Psi^T_{\beta}$ denote the
wave functions of a projectile nucleus $P$ and a target nucleus $T$. 
The cross section from  an initial state ($\alpha=0, \beta=0$)
to a final state, $\alpha$ and $\beta$, is evaluated by 
integrating the reaction probability over an impact parameter vector $\bm {b}$~\cite{Glauber}:
\begin{align}
  \sigma_{\alpha\beta}
=\int d\bm{b}\,
  |\left<\Psi^P_\alpha\Psi^T_\beta\right|
  \prod_{j\in P}\prod_{k \in T}[1-\Gamma_{jk}]
  \left|\Psi^P_0\Psi^T_0\right>|^2.  
\end{align}
Here, $\Gamma_{jk}= 1- e^{i\chi_{NN}(\bm{b}+\bm{s}_j-\bm{s}_k)}$
is specified by a phase-shift function $\chi_{NN}$ of the nucleon-nucleon ($NN$) scattering. 
$\bm{s}_j$ (${\bm s}_k$) is a two-dimensional vector, perpendicular to the beam ($z$) direction, of the nucleon coordinate relative to the center of mass of 
the projectile (target) nucleus. 
$\chi_{NN}$ depends on $pn$ or 
$pp$ pair. $\chi_{nn}$ is assumed to be the same as $\chi_{pp}$.
Spin-dependence of $\chi_{NN}$ is ignored.  
The parameters of $\Gamma_{NN}$ are taken from Ref.~\cite{Ibrahim08}.

The total reaction cross section is defined by 
$\sigma_{T:R}=\sum_{\alpha \beta}\sigma_{\alpha \beta}-\sigma_{00}$.
The sum over $\alpha \beta$ is taken by using a closure relation,
e.g., $\sum_\alpha \left|\Psi^P_\alpha\right>\left<\Psi^P_\alpha\right|=1$,
and the unitarity of the $NN$ phase-shift function, 
$|e^{i\chi_{NN}}|^2=1$, leading to 
\begin{align}
  \sigma_{T:R}=\int d{\bm b} \, (1-P_{PT}(\bm b)),
\label{reac.eq}
\end{align}
where $P_{PT}(\bm b)$ is the probability for the elastic scattering 
\begin{align}
P_{PT}(\bm b)=\Big|\langle\Psi^P_0\Psi^T_0|\prod_{j\in P}\prod_{k \in T}[1-\Gamma_{jk}] |\Psi^P_0\Psi^T_0\rangle\Big|^2.
\label{prob.elastic}
\end{align}

Although the evaluation of $P_{PT}(\bm b)$ is in general hard,
a nucleus-$N$ case can be done to good accuracy.
Given the proton and neutron densities of the projectile nucleus $P$, we get 
the $PN$ phase-shift function $\chi^P_{N}$ by~\cite{Glauber,Suzuki03}
\begin{align}
  &i\chi^P_{N}(\bm{b})=\ln \left<\Psi^P_0\right|
  \prod_{j\in P}[1-\Gamma_{jN}]
  \left|\Psi^P_0\right>\notag\\
  &\approx-\int d\bm{r}\,
  \left[\rho^{P}_p(\bm{r})\Gamma_{pN}(\bm{b}+\bm{s})+
    \rho^{P}_n(\bm{r})\Gamma_{nN}(\bm{b}+\bm{s})\right],
\label{OLA.eq}
\end{align}
where $\bm r=(\bm s, z)$. This approximation works well
when the fluctuation of $\chi_N^P$
is small enough~\cite{Yabana92a, Yabana92b, Ogawa92, Suzuki03}.
In fact, the ansatz~(\ref{OLA.eq})
that relates $\chi^P_N$ to the proton and neutron densities of $P$ 
has proven to work well for many cases of $PN$
scattering~\cite{Varga02, Ibrahim09, Nagahisa18, Hatakeyama19}.
The $PN$ total reaction 
cross section $\sigma_N$ ($N=n, p$) reads 
\begin{align}
  \sigma_{N}=\int d\bm{b}\ (1-P_N(\bm{b}))
\end{align}
with
\begin{align}
  P_N(\bm{b})=\big|e^{i\chi^P_{N}(\bm{b})}\big|^2.
\end{align}

A unique advantage of the deuteron target is that we can calculate 
Eq.~(\ref{prob.elastic}) accurately using its ground-state wave function $\phi_d(\bm{r})$. The $Pd$ total reaction cross section turns out to be 
\begin{align}
  \sigma_d=\int d\bm{b}  \left(1-P_d(\bm b)\right),
\label{sigr.eq}
\end{align}
where
\begin{align}
P_d(\bm b)=\Big|\int d\bm{r}\,|\phi_d(\bm{r})|^2
  e^{i\chi^P_{p}(\bm{b}+\frac{1}{2}\bm{s})+i\chi^P_{n}(\bm{b}-\frac{1}{2}\bm{s})}\Big|^2.
  \label{ga1.eq}
\end{align}
We use the AV8$^\prime$ potential~\cite{AV8} to generate $\phi_d(\bm{r})$.  
$|\phi_d(\bm{r})|^2$ is actually defined  by taking an average with  respect to relevant magnetic quantum numbers and by integrating out  
spin-isospin coordinates. 
In this way we get a quantitative evaluation of the cross-section defect: 
\begin{align}
  \delta \sigma=\sigma_p+\sigma_n-\sigma_d.
\label{crosssection.defect}
\end{align}

\subsection{Interaction cross section}

Experimentally observed in most cases is the interaction cross section 
but not the total reaction cross section. Since it is hard to calculate 
the interaction cross section at the same accuracy as the total reaction 
cross section, it is of practical importance 
to set a theoretical limit on the interaction cross section. 
The interaction cross section is a semi-inclusive 
cross section defined by summing over all $\alpha$'s but
particle-bound states (b.s.): 
\begin{align}
\sigma_{T:I}=\sum_{\alpha\beta}\sigma_{\alpha\beta}
  -\sum_{\alpha \in {\rm b.s.}\, \beta}\sigma_{\alpha\beta}
  =\sigma_{T:R}-\Delta\sigma,
\end{align} 
where
\begin{align}
\Delta \sigma=\sum_{\alpha \in {\rm b.s.}\, \beta}\sigma_{\alpha\beta} - \sigma_{00}.
\end{align}

A calculation of $\Delta\sigma$ demands all b.s. wave functions
of the projectile nucleus, which is in general hopeless. 
Provided the projectile nucleus has only one b.s., its ground state, 
$\sigma_{T:I}$ is equal to $\sigma_{T:R}-\Delta_0 \sigma$, where  
\begin{align}
  \Delta_0\sigma = \sum_{\beta}\sigma_{0\beta}-\sigma_{00}.
\label{del0sig}
\end{align}
In other cases $\Delta_0 \sigma$ is the lower bound of 
$\Delta \sigma$: $\Delta_0 \sigma \leqq \Delta \sigma$. Even the calculation of $\Delta_0 \sigma$ is difficult in a general case.
For the deuteron target, however, with 
the same ingredients as needed to calculate
$\sigma_d$, we can get it. To show this,
  we again use the closure 
  relation for all target states and the approximation of Eq.~(\ref{OLA.eq})
  to the phase-shift  function: 
\begin{align}
  &\sum_{\beta}\sigma_{0\beta}\notag\\
  &=\int d\bm{b}\,
  \left<\Psi^T_0\right|\big|
  \left<\Psi^P_0\right|\prod_{j\in P}\prod_{k \in T}[1-\Gamma_{jk}]
  \left|\Psi^P_0\right>
  \big|^2 \left|\Psi^T_0\right>\\
  &\simeq\int d \bm{b}\, P_0(\bm b),
\end{align}
where 
\begin{align}
P_0(\bm b)=\int d\bm{r}\,|\phi_d(\bm{r})|^2
      \left|e^{i\chi^P_{p}(\bm{b}+\frac{1}{2}\bm{s})+i\chi^P_{n}(\bm{b}-\frac{1}{2}\bm{s})}\right|^2.
\label{ga2.eq}
\end{align}
Substitution of this result into Eq.~(\ref{del0sig})
leads to
\begin{align}
&\Delta_0 \sigma=\int d\bm{b} \left(P_0(\bm b)-P_d(\bm b)\right).
\label{sigi0.eq}
\end{align}

\section{Results and discussion}
\label{results.sec}

\subsection{Test of  $\sigma_d$ calculation}
\label{test.sec}

First, we compare in Table~\ref{sigr.tab} the $\sigma_d$ values
with experiment for some well-known nuclei, $^{12}$C, $^{16}$O, and $^{40}$Ca.
The values in parentheses denote
the results with the Coulomb breakup cross sections.
The center-of-mass-corrected harmonic-oscillator type
densities  are employed for those nuclei~\cite{Ibrahim09}, and 
the oscillator parameters needed to specify the densities are set 
to reproduce empirical charge radii~\cite{Angeli13}.
Although the data are at low incident energies,
the Glauber-model calculation reproduces them very well.
The inclusion of the Coulomb breakup contribution leads to  
results further closer to experiment especially for $^{40}$Ca. 
Our $\sigma_d$ value for $^{40}$Ca is in excellent agreement 
with  1270 mb, which is the cross section obtained by a continuum-discretized-coupled-channels  calculation~\cite{Minomo17} that includes the Coulomb breakup.  

The table also lists the cross sections calculated with an optical-limit approximation (OLA)~\cite{Glauber,Suzuki03}, which uses 
\begin{align}
P_d^{\rm OLA}(\bm b)
=\big|e^{ i  \int d\bm{r}\,
    [\rho^d_{p}(\bm{r})\chi^P_{p}
    (\bm{b} +\bm{s})+\rho^d_{n}(\bm{r})\chi^P_{n}
    (\bm{b} +\bm{s})]
}\big|^2,
\label{OLAd.eq}
\end{align}
instead of Eq.~(\ref{ga1.eq}), where $\rho^d_{N}$ is the nucleon-density of the deuteron. Since the OLA takes only the first term
in the cumulant expansion, this approximation
is not expected to work well in the $Pd$
scattering, where the deuteron wave function is spatially extended
~\cite{Yabana92a, Yabana92b, Ogawa92}.
Apparently the OLA tends to overestimate $\sigma_d$ significantly. 

\begin{table}[ht]
  \begin{center}
    \caption{Total reaction cross sections of
      $^{12}$C, $^{16}$O, and $^{40}$Ca on a deuteron target
      at 50 MeV/nucleon. 
    The experimental cross sections at 48.7$\pm 0.15$ MeV/nucleon
    are taken from Ref.~\cite{Auce96}. The 
    cross section in parenthesis 
    includes the Coulomb breakup contribution of deuteron that is 
    estimated following Ref.~\cite{Horiuchi16}.}
  \label{sigr.tab}
    \begin{tabular}{cccccc}
        \hline\hline
        && Glauber&OLA&Expt.~\cite{Auce96} \\
        \hline                             
        $^{12}$C&& 638 (642)&  687& $600\pm 17$  \\
        $^{16}$O&& 735 (741)&  795 & $726\pm 21$   \\
        $^{40}$Ca&&1220 (1260)&1310 & $1260\pm 30$  \\
        \hline\hline
    \end{tabular}
    \end{center}
\end{table}

In what follows we neglect the Coulomb breakup contribution, although  
its inclusion is possible in the Glauber model as 
discussed in Refs.~\cite{Margueron03,Ibrahim04,Capel08,Horiuchi16,Horiuchi17}.  
It should be noted that the most important Coulomb effect for heavy projectiles is the deuteron breakup: It increases $\sigma_d$ by 
a few percent for the projectile nucleus with its proton number $Z_P\approx 20$ 
and by about $5$ percent for $Z_P\approx 40$~\cite{Horiuchi16}. 

\begin{figure}[ht]
\begin{center}
 \epsfig{file=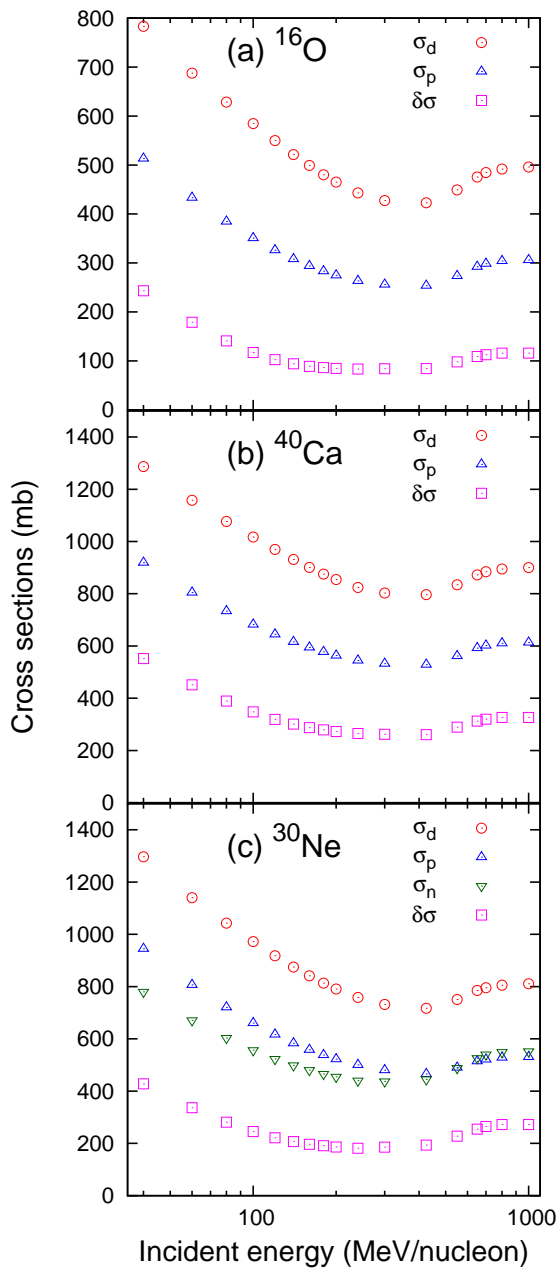, scale=1.2}
 \caption{Total reaction cross sections,  $\sigma_d$, $\sigma_p$, and $\sigma_n$ of the projectile nuclei of (a) $^{16}$O,  (b) $^{40}$Ca, and  (c) $^{30}$Ne 
as a function of incident energy. $\sigma_n$ is equal to 
$\sigma_p$ for $^{16}$O and $^{40}$Ca.  
The cross-section defect $\delta \sigma$  is also plotted.  }
\label{rcs.fig}
\end{center}
\end{figure}

Figure~\ref{rcs.fig} displays $\sigma_d$,  $\sigma_p$, $\sigma_n$  
as well as $\delta \sigma$ for (a) 
$^{16}$O, (b) $^{40}$Ca, and (c) $^{30}$Ne 
as a function of incident energy. Note that $\sigma_n$ is equal to 
$\sigma_p$ for $^{16}$O and 
$^{40}$Ca because the neutron and proton densities are identical. 
The density distribution of $^{30}$Ne, taken from
Refs.~\cite{Horiuchi12, Horiuchi14},  is obtained by 
the Skyrme-Hartree-Fock calculation with
the SkM$^*$ effective interaction~\cite{SkMs}. Note that 
the density of $^{30}$Ne reproduces reasonably well 
the observed cross section
on a carbon target at 240 MeV/nucleon~\cite{Takechi10}. 
The $\sigma_p$ value of $^{30}$Ne is close to 
that of $^{40}$Ca, which is due to the fact that $\sigma_p$ is most 
sensitive to the neutron density of the projectile nucleus and both $^{30}$Ne and $^{40}$Ca have almost the same matter radius. 
In contrast to $\sigma_p$, 
$\sigma_n$ for $^{30}$Ne is smaller than that for $^{40}$Ca as understood 
from different proton numbers of those projectile nuclei.  

For the above three projectile nuclei, $\sigma_d$ is always significantly 
smaller than $\sigma_p+\sigma_n$: The $\delta\sigma$ value is about 
$ 10$--30\% of $\sigma_p+\sigma_n$.
The larger mass number of the projectile nucleus, 
the larger $\delta\sigma$ value is obtained,  
which is naturally understood from the eclipse effect:  
The chance that the proton and neutron in the deuteron cast individual shadows 
increases with increasing mass number of the projectile nucleus.
The energy dependence of $\delta \sigma$ follows that of the $NN$ 
total cross section $\sigma_{NN}^{\rm tot}$ that specifies $\Gamma_{NN}$ 
(see, e.g., Ref.~\cite{PDG}). $\delta \sigma$ 
reaches a minimum at $\approx 200$--500 MeV/nucleon,
where $\sigma_{NN}^{\rm tot}$ also becomes a minimum.

A simple estimate of $\delta \sigma$ for $\bar{p}d$ case 
was given by 
$\delta\sigma=2\sigma_p \sigma_n\left<\frac{1}{4\pi r^2}\right>_d$~\cite{Glauber, Franco66}, 
where $\left< \frac{1}{r^2} \right>_d$ stands for 
the expectation value of the inverse square of the neutron-proton distance $r$
  by the deuteron ground state. With our present 
value, 
$\left<\frac{1}{r^2}\right>_d=0.294$ fm$^{-2}=0.0294$ mb$^{-1}$, we find that 
  the formula significantly
  overestimates the cross-section defect, about 5 times larger
  than the present one. The formula was actually derived under the 
 assumption that 
  the size of the deuteron is larger than 
  the range of the $\bar{p}N$ interaction. The assumption is not valid 
 in our case because the range of the $PN$ interaction is at most 
 comparable or larger than the size of the deuteron.

\subsection{Numerical experiment: extracting $\sigma_n$}
\label{extract.sec}

Suppose that both $\sigma_p$ and $\sigma_d$ are measured experimentally. 
An interesting 
question is whether or not we can extract $\sigma_n$ from those observed cross 
sections. 
To discuss its possibility, 
we perform a numerical experiment by taking examples of 
$N_P=2Z_P$ projectile nucleus, $^{30}$Ne and $^{60}$Ca, where 
the difference between $\sigma_p$ and $\sigma_n$ is expected to be 
large especially at low incident energy because of  
large neutron skin thickness expected. 

We assume $\sigma_p$ and $\sigma_d$ calculated with 
the HF density distribution to be experimental data, denoted
as $\sigma_p^{\rm HF}$ and $\sigma_d^{\rm HF}$, respectively. 
We attempt to determine $\sigma_n$ under the assumption that 
no knowledge on the density distribution or even the radius of $P$ is given. 
In any case we need the density of $P$, and as a reasonable choice 
we assume two-parameter 
Fermi (2pF) function for neutron and proton ($N=n, p$) 
\begin{align}
\rho_{N}(r)=\frac{\rho_{0N}}{1+\exp\left[(r-R_N)/d_N\right]},
\label{2pF.eq}
\end{align}
where $R_N$ and $d_N$ are radius and diffuseness parameters,
and the central density $\rho_{0N}$ is determined
by the condition, e.g., for proton $4\pi \int_0^{\infty}dr r^2 \rho_p(r)=Z_P$. 
We consider two cases below.

\subsubsection*{Case I: Data available at two incident energies}

We may assume different parameters for the Fermi functions of 
neutron and proton, and determine them so as to reproduce the  
`observed' $\sigma_d^{\rm HF}$ and $\sigma_p^{\rm HF}$ values at two 
incident energies, chosen at 100 and 200 MeV/nucleon. Whether this procedure 
is successful or not is judged by comparing resulting `theoretical' values of 
$\sigma_n$ and $\delta\sigma$ with $\sigma_n^{\rm HF}$ and $\delta\sigma^{\rm HF}$. 
Plots labeled 2pF1 in Fig.~\ref{rcsN2Z.fig} denote the theoretical values for 
the incident energy of 40 to 1000 MeV/nucleon, 
whereas  plots labeled 
HF observed ones.  Both cases of (a) $^{30}$Ne and (b) $^{60}$Ca projectile 
nuclei show excellent agreement: The 2pF density distributions 
reproduce the HF cross sections of $^{30}$Ne
and $^{60}$Ca for {\it all} the incident energies
in spite of the fact that 
the four parameters of the Fermi functions are determined 
to reproduce the experimental data only at two incident energies. 
This numerical experiment strongly corroborates that  
the evaluation of $\sigma_d$ is reliable enough to predict $\sigma_n$ correctly 
even at different incident energies.

The extracted parameters, ($R_n, d_n$) and ($R_p, d_p$), are, in fm, 
(3.20, 0.66) and (3.08, 0.50) for $^{30}$Ne, 
and (4.33, 0.65) and (4.03, 0.52) for $^{60}$Ca, respectively. 
We see different surface diffuseness for neutron and proton
and large neutron skin thickness for both nuclei. Let us define the 
matter radius $r_m$ and neutron skin thickness $\delta r$ of $P$ in terms of 
the root-mean square radii of neutron and proton, $r_n(N_P,Z_P)$ and 
$r_p(N_P,Z_P)$, by 
\begin{align}
&r_m=\sqrt{\frac{Z_P}{A_P}r_p^2(N_P,Z_P)+\frac{N_P}{A_P}r_n^2(N_P,Z_P)},\notag \\
&\delta r=r_n(N_P,Z_P)-r_p(N_P,Z_P),
\end{align}
where $A_P=N_P+Z_P$. The resulting values 
are $(r_m,\delta r)=(3.34, 0.47)$ for $^{30}$Ne
and (3.98, 0.46) for $^{60}$Ca, in excellent agreement
with the HF values (3.34, 0.47) and (4.00, 0.46), respectively. 

In order for the present analysis to be sensitive enough to the choice of the 
parameters of the Fermi function, one of the incident energies should be chosen 
$\lesssim 300$ MeV/nucleon,  because 
$\sigma_{pn}^{\rm tot}$ is then much larger than $\sigma_{pp}^{\rm tot}$ and  
reduces possible uncertainty in determining the parameters~\cite{Horiuchi14,Horiuchi16}. 

\subsubsection*{Case II: Data available at only one incident energy}

\begin{figure}[ht]
\begin{center}
 \epsfig{file=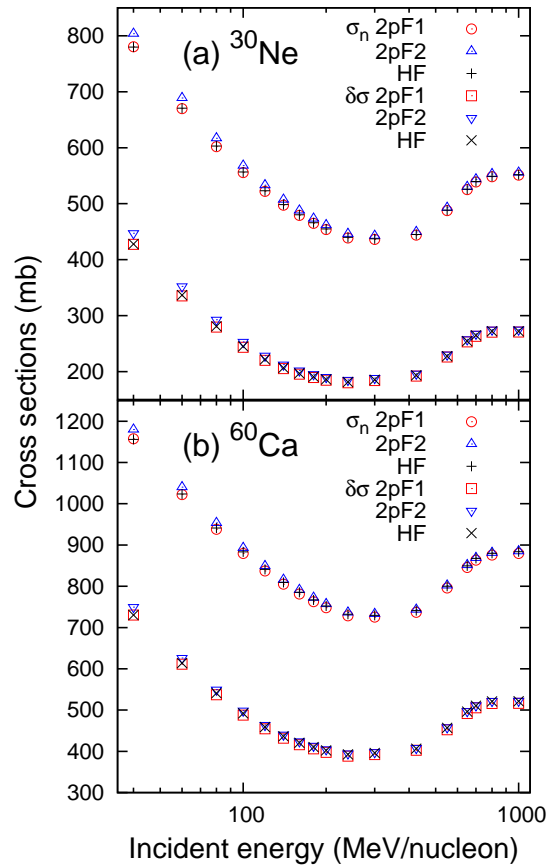, scale=1.2}
 \caption{Comparison between `theory' (2pF1 and 2pF2) and `experiment' (HF) for 
total reaction cross section $\sigma_n$ of 
  (a) $^{30}$Ne and (b) $^{60}$Ca and cross-section defect 
$\delta \sigma$. 2pF1 (2pF2) stands for the cross sections determined so as to 
reproduce $\sigma_p$ and $\sigma_d$  
   values at selected incident energies, assuming different (conditional) Fermi-type density 
 distribution  for neutron and proton.  See text for detail.} 
\label{rcsN2Z.fig}
\end{center}
\end{figure}

If we have experimental data,  $\sigma_p$ and $\sigma_d$,
at only one incident energy, say 
100 MeV/nucleon, we put a constraint, e.g., $d_n=d_p=0.6$ fm and attempt to  
reproduce the experimental data using two free parameters, $R_n$ and $R_p$. 
The extracted radius parameters, $(R_n, R_p)$,  are in fm   
(3.42, 2.90) $^{30}$Ne, and (4.49, 3.88) for $^{60}$Ca,  respectively.
Though the matter radii $r_m$ of these isotopes
are 3.37 and 4.01 in fair agreement with the HF values,
the resulting $\delta r$ values turn out to be 
0.30 for $^{30}$Ne and 0.39 for 
$^{60}$Ca, which are considerably smaller than the HF values. 
As shown in Fig.~\ref{rcsN2Z.fig},
the calculated $\sigma_n$ value denoted 2pF2 still reproduces the HF value 
reasonably well at higher incident energies but 
deviates from $\sigma^{\rm HF}_n$ with decreasing incident energies, e.g., 
by about  2\% at 40 MeV/nucleon. 
The quality of the theoretical prediction of course depends on the incident 
energy chosen for the data, and it should be as low as $\lesssim 300$ 
MeV/nucleon.
\\

\begin{figure*}[ht]
\begin{center}
      \epsfig{file=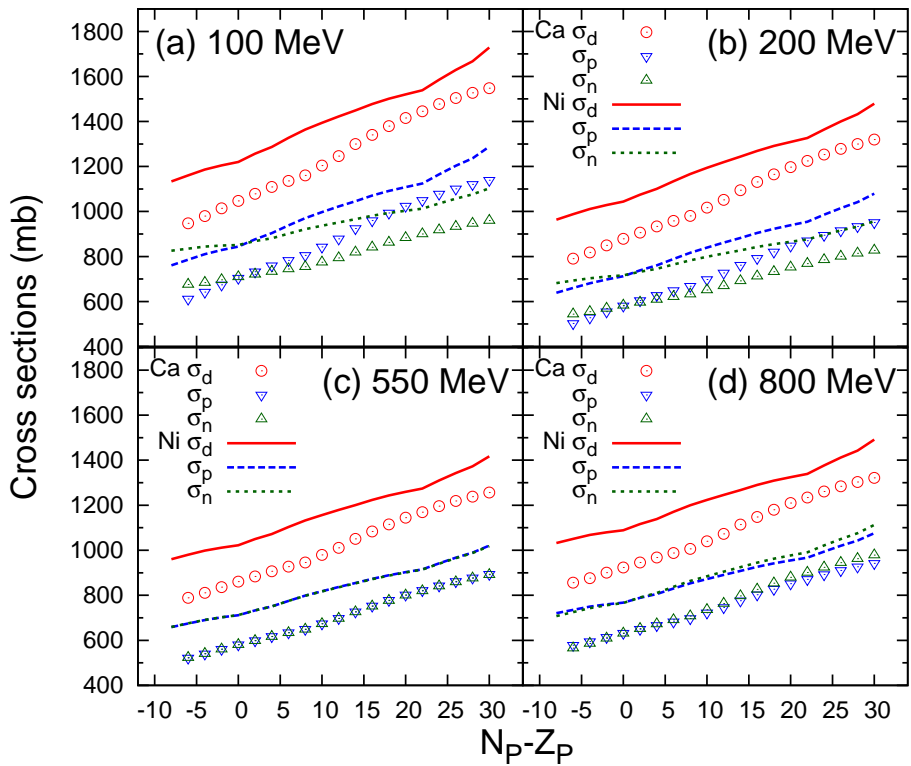, scale=1.3}
      \caption{Total reaction cross sections of $\sigma_d$, $\sigma_p$, and $\sigma_n$ for Ca and Ni isotopes at the incident energy of (a) 100, (b) 200, (c) 550, and (d) 800 MeV/nucleon 
        as a function of neutron excess $N_P-Z_P$.}
\label{HF_rcs.fig}
\end{center}
\end{figure*}

As confirmed in the Case I study, we have succeeded to calculate consistently 
$\sigma_d$, $\sigma_p$, and $\sigma_n$ cross sections at all incident energies. 
In Fig.~\ref{HF_rcs.fig} we exhibit those cross sections for $^{34-70}$Ca and 
$^{48-86}$Ni 
as a function of neutron excess $N_P-Z_P$. The densities 
of Ca and Ni isotopes are based on HF 
calculation~\cite{Horiuchi12,Horiuchi14}. In so far as the HF calculation  gives 
$r_m$ and $\delta r$ close to realistic values, the cross sections in the figure are expected 
to show general feature of real cross sections. For example, 
the cross section ratio, $\sigma_p / \sigma_n$, shows an interesting 
variation against the incident energy $E$: The ratio is larger than one 
for $N_P-Z_P>0$ and smaller than one for $N_P-Z_P<0$ at both $E=100$ and 200 MeV/nucleon, 
whereas it is almost 1 at $E=550$ MeV/nucleon, and gets smaller than 1 at 
800 MeV/nucleon. 
This variation of the ratio exactly follows that of 
$\sigma_{pn}^{\rm tot}/\sigma_{pp}^{\rm tot}$. 

  We note, however, that the uncertainties of the present $\sigma_d$ data
  are approximately 2--3\%~\cite{Auce96}, which are the same
  order or even larger than the deviation obtained in the Case II study.
  Improvement of the experimental accuracy
  is highly desired to precise determination
  of $\sigma_n$ as it was already achieved $\lesssim 1$\%
  for the interaction cross section measurements involving
  unstable nuclei by
  proton~\cite{Moriguchi20} and carbon targets~\cite{Tanaka20}.

\subsection{Relating  $\sigma_{T:R}$ to $r_m$ and $\delta r$}
\label{emp.sec}

The numerical analysis of Case I has confirmed that 
$\sigma_n$ or even  $(r_m, \delta r)$ values of a projectile nucleus $P$ 
can be reproduced quite well once $\sigma_d$ and $\sigma_p$ of $P$ 
are given. This strongly suggests that the total 
reaction cross section $\sigma_{T:R}$ of $P$ can be determined by $r_m$ and $\delta r$ to 
good accuracy: Introducing a reaction radius $a_T$ by $\sigma_{T:R}=\pi a_T^2$, we may conjecture that $a_T$ is expressed as 
\begin{align}
  a_T=\alpha_T(E)r_m+\beta_T(E)\delta r +\gamma_T(E).
\label{reaction.radius}
\end{align}
Here, $\alpha_T(E), \ \beta_T(E)$, and $\gamma_T(E)$ depend on the incident 
energy $E$ but do not depend on the projectile nucleus. 
The conjecture has actually been confirmed successfully 
in Refs.~\cite{Horiuchi14, Horiuchi16} for proton and $^{12}$C targets. 
We extend that analysis to determine the coefficients for the 
deuteron and neutron targets by covering 
the projectile nuclei, $^{14-24}$O, $^{18-34}$Ne, $^{20-40}$Mg,
$^{22-46}$Si, $^{26-50}$S, $^{34-70}$Ca, and $^{48-86}$Ni. See 
Refs.~\cite{Horiuchi14, Horiuchi16} for detail.

\begin{figure}[ht]
\begin{center}
     \epsfig{file=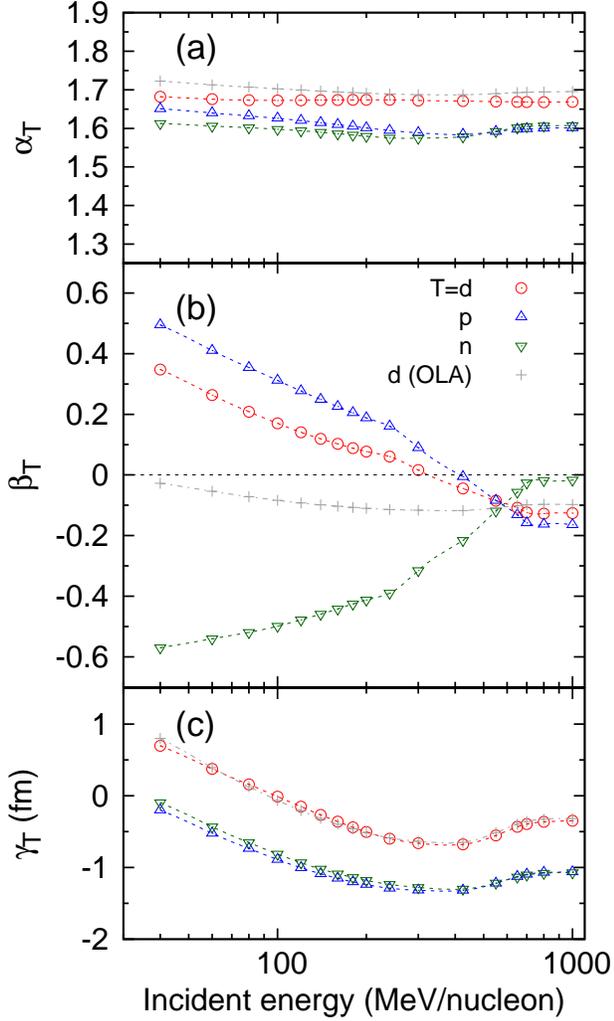, scale=1.3}
      \caption{Incident-energy dependence of coefficients (a) 
        $\alpha_T$, (b) $\beta_T$, and (c) $\gamma_T$ of the reaction radius 
        $a_T$ of Eq.~(\ref{reaction.radius}), where $T$ denotes a target 
nucleus.   d(OLA) stands for the coefficients obtained when $\sigma_d$ is 
calculated by the OLA. }
\label{ecoef.fig}
\end{center}
\end{figure}

Figure~\ref{ecoef.fig} compares those coefficients
for deuteron, proton, and neutron targets. 
$\alpha_T$ depends on $E$ weakly, which is understood by noting that 
$\sigma_{T:R}$ is roughly equal to  $\pi r_m^2$. 
The coefficient $\beta_T$ among others is most important to probe 
$\delta r$. The larger $\beta_T$ is in its magnitude, the more accurately 
$\delta r$ is determined. The $E$-dependence of $\beta_T$ follows that of 
$\sigma_{NN}^{\rm tot}$: The ratio $\sigma_{pn}^{\rm tot}/\sigma_{pp}^{\rm tot}$
is about 2.5 at 100 MeV/nucleon, 2 at 200 MeV/nucleon, about 1 
at $\approx$550 MeV/nucleon, and less than 1 beyond 800 MeV/nucleon. 
The apparent symmetry of $\beta_n$ and $\beta_p$ is understood by noting 
that $\sigma_n$ can also be obtained from $\chi^P_p(\bm b)$ of 
Eq.~(\ref{OLA.eq}) with $\rho^P_p(\bm r)$ and $\rho^P_n(\bm r)$ being 
exchanged. 

What is noteworthy in the figure is that $\beta_d$ has fairly large 
values especially at lower energies, which is different from the $^{12}$C target~\cite{Horiuchi14}. The larger sensitivity of the deuteron target is probably 
because the deuteron is spatially extended and may probe sensitively the 
surface region of the projectile nucleus. 
If $\sigma_d$ is, however, calculated by OLA, 
Eq.~(\ref{OLAd.eq}), the resulting coefficients differ from those discussed above. The coefficients given in the OLA calculation are also displayed in the figure. We find that $\beta_d$ with OLA is found to be small and 
almost constant $\approx -0.1$ beyond 100 MeV/nucleon. It is better to avoid 
OLA calculations for an analysis with $\sigma_d$ data.

\subsection{Difference between $\sigma_d$ and $\sigma_{d:I}$}
\label{interaction.sec}

\begin{figure}[ht]
\begin{center}
\epsfig{file=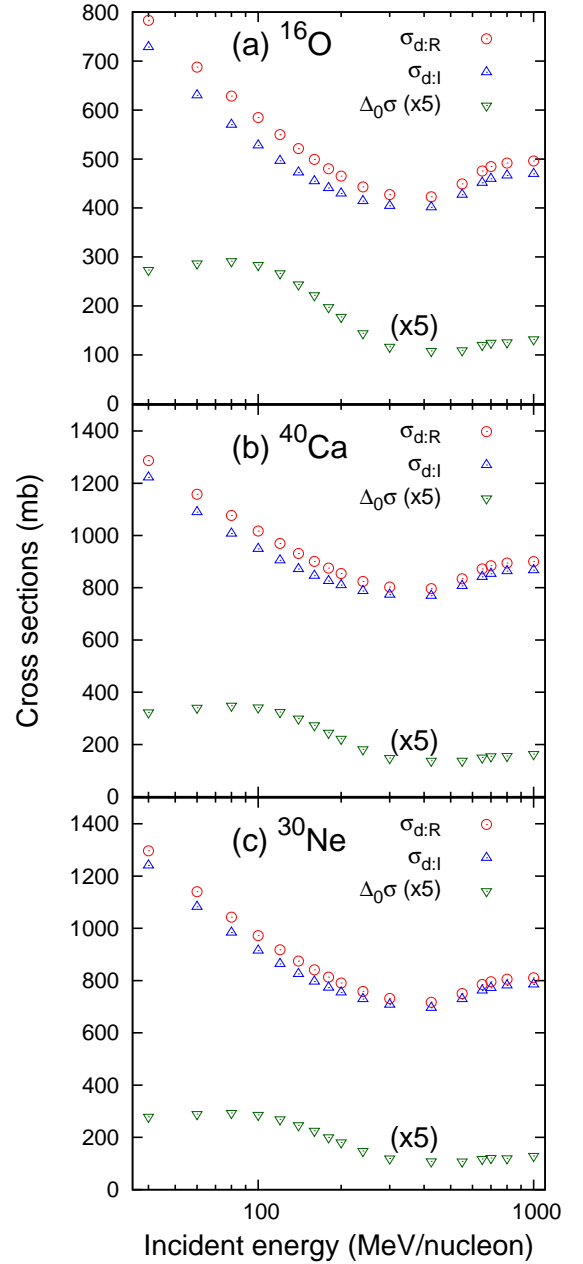, scale=1.2}
 \caption{
    Total reaction cross section $\sigma_d$ and  interaction cross section $\sigma_{d:I}$ for 
     (a) $^{16}$O, (b) $^{40}$Ca, and (c) $^{30}$Ne projectile nucleus 
   as a function of incident energy.   $\sigma_{d:I}$ shown here is actually $\sigma_d-\Delta_0 \sigma$, 
which is equal to the upper bound of $\sigma_{d:I}$.   $\Delta_0\sigma$  multiplied
   by 5 is also plotted.}
\label{sigi.fig}
\end{center}
\end{figure}

As already mentioned, the deuteron target has the advantages that 
the upper bound of the interaction cross section can be evaluated 
reliably using the deuteron wave function, that is, $\sigma_{d:I} \leqq \sigma_d-\Delta_0 \sigma$, 
where  $\Delta_0 \sigma$ is calculated by using  Eq.~(\ref{sigi0.eq}). 
Figure~\ref{sigi.fig} displays $\sigma_d$,  $\sigma_d-\Delta_0 \sigma$\, (the upper bound of $\sigma_{d:I}$), and $\Delta_0 \sigma$ for (a) $^{16}$O, (b) $^{40}$Ca and (c) $^{30}$Ne  projectile 
nucleus as a function of incident energy.  $\Delta_0 \sigma$ 
has a maximum at 80 MeV/nucleon and its magnitude is 60--70 mb, 
which is about 10\% of $\sigma_d$ for $^{16}$O, 7\% for $^{40}$Ca, and 6\% for 
$^{30}$Ne, respectively.  
$\Delta_0\sigma$ decreases with increasing incident energy and 
has a minimum at 425 MeV/nucleon. It looks that the ratio 
$\Delta_0\sigma /\sigma_d$ has a minimum at 550 MeV/nucleon. Beyond 300 
MeV/nucleon, the ratio is at most few percent for both $^{40}$Ca and $^{30}$Ne.

\section{Conclusion}
\label{conclusion.sec}

We have systematically investigated the total reaction
cross sections scattered by a deuteron target
using the Glauber model. The eclipse effect cast by the neutron and 
proton in the deuteron is explicitly taken into account.
The interaction cross section is also evaluated
in the same framework.

The calculated total reaction cross sections reasonably well 
reproduce the available experimental data. By extending the analysis 
to cover many nuclei, O to Ni,  we find that  
the nucleus-deuteron cross section is considerably smaller than the 
sum of cross sections of nucleus-neutron and nucleus-proton. The 
cross-section defect is understood by the eclipse effect.  

The cross-section defect carries information on 
the nucleus-proton and nucleus-neutron profile functions. 
Because of this, the nucleus-neutron cross section can be extracted 
by simultaneous measurements of the total reaction cross sections 
by both deuteron and proton targets. 
We have convincingly shown that measuring the both cross sections  
at two incident energies is the best choice to determine the neutron 
cross section or the nuclear size properties. 
Energy-dependence of the total reaction cross section 
is given in terms of the matter radius and neutron skin thickness of the 
projectile nucleus. 

Measuring the total reaction cross sections
by both deuteron and proton targets is the most unambiguous way
to determine the neutron and proton radii of unstable nuclei.
We note that for the unstable nuclei near the dripline
that have only one bound state,
the theoretical interaction cross section
can be obtained in good accuracy.
In the present analysis, we ignore the Coulomb breakup
contribution of the deuteron target,
which will be significant for heavy projectiles.
The inclusion of this effect to the Glauber model
is straightforward and will be reported elsewhere.

\acknowledgments

We thank K. Ogata for valuable communications.
This work was in part supported by JSPS KAKENHI Grants
Nos. 18K03635, 18H04569, and 19H05140.
WH acknowledges the Collaborative Research Program 2020, 
Information Initiative Center, Hokkaido University.

\end{document}